# Multidecadal Cycles of the Climatic Index Atlantic Meridional Mode: Sunspots that Affect North and Northeast of Brazil

Cleber Souza Correa[1,*], Roberto Lage Guedes[1], André Muniz Marinho da Rocha[1], Karlmer Abel Bueno Corrêa[2]

Correa CS  https://orcid.org/0000-0001-5799-1982
Guedes RL  https://orcid.org/0000-0002-7886-5974
da Rocha AMM  https://orcid.org/0000-0003-3611-781X
Corrêa KAB  https://orcid.org/0000-0001-9078-1514

**ABSTRACT:** Using the 1951-2017 historical series of the Atlantic Meridional Mode (AMM) index and the monthly number of sunspots, it was possible to observe a significant association between them. The use of wavelet and cross-wavelet analysis showed the presence of multidecadal cycles pronounced in eleven years, as well as cycles of 2.66 and 5.33. AMM index showed, in the part of the Sea Surface Temperature (SST), the presence of a weak signal of 21.33 years. Influence and association of sunspot variability on surface temperature in Northern and Northeastern regions of Brazil were investigated. Using a non-parametric statistical correlation test, the historical series of surface temperature anomalies in five locations (Belém, São Luiz, Fortaleza, Fernando de Noronha, and Natal) were compared with the monthly solar-series anomalies. The temperature series used were the minimum monthly average, the monthly average, and maximum monthly average temperatures, with their respective anomalies in relation to the mean. However, among all the series (except for São Luiz), the analyzed minimum temperature anomalies showed a negative correlation with sunspots. As a preliminary result, the analyzed climatic indexes present an apparent degree of memory associated with the variability of sunspot activity.

## INTRODUCTION

This work focuses on analyzing the temporal sunspot series and the AMM SST index information, in order to observe similar cycles between time series, since there is a gap between the influences of solar activity on terrestrial systems, such as in atmospheric and oceanic dynamics. This is associated with an interhemispheric gradient of anomalous SST, in which part of the dominant frequencies of turbulent processes is known at different associated timescales, varying from monthly, annual and multidecadal. It can, however, also be influenced by teleconnection processes between the planetary regions (*e. g.*, the dynamics of the El Niño-Southern Oscillation [ENSO] circulation).

In Corrêa *et al.* (2019), using wavelet and cross-wavelet analysis, multidecadal cycles were observed between the monthly number of spots and the Southern Oscillation Index (SOI) and Pacific Decadal Oscillation (DOP) indexes, showing cycles of 2.66, 5.33, 10.66 and 21.33 years. It was also compared to the average monthly rainfall in the meteorological stations of the airports of Belém, Fortaleza, São Luiz, and Natal, showing that, in North/Northeast of Brazil, the multidecadal cycles of precipitation accompanied the variability of the sunspots.

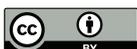

Prestes *et al.* (2018) showed that the araucaria tree responds to solar activity, variability, and climate in Southern Brazil. Using wavelet analysis, the researchers found periodicities in the rings of araucaria trees growth, associated with the SOI, as well as with an anomaly of the average annual temperature, between 24 and 44 ºC. It was also observed that there were periods of 2 to 7 years, possibly related to the ENSO phenomenon, and a second period, which lasted for approximately 23 years and was related to temperature variation.

Cliver (2015) discusses the cycles of solar activity and the 22-year magnetic cycle of the Sun, in which there are two characteristic migrations in the behavior of sunspots: the movement towards the solar equator and the movement towards the poles of prominence at high latitudes. The motion of spots towards the equator are defined aspects associated with Schwabe's eleven-year cycles (Schwabe 1843; Wilson 1998) and the second is associated with the 22-year magnetic cycles, otherwise known as Hale's solar cycle (22 years) (Hale and Nicholsen 1925; Babcock 1961; Echer *et al.* 2003).

Lassen and Friis-Christensen (1995) showed that the duration of a solar cycle in the last five centuries was associated with Earth's climate, having a well-defined activity with an eleven-year cycle in the sunspots number. Solar activity data have been used as parameters for studies addressing physical conditions variability of the high Earth atmosphere. Solar activity can vary in time, with cycles of eleven years. In addition, the duration of sunspots periods is also not fixed, varying between 80 and 90 years. Strong correlations have been observed between these changes and the long-term variations in global temperature.

There are studies that make the association between solar variability and Earth climate (Cliver *et al.* 1998; Reid 2000; Love *et al.* 2011; Wang *et al.* 2018), in which solar activity, such as the average number of annual sunspots and the mean global SST, and the average global temperature are associated. The changes related to the intensity of solar activity on planet Earth's temperature, and because it has different surfaces (continental and oceanic), create a global circulation, associated with the dynamic processes between continents and the oceanic surfaces. The global circulation creates turbulent dynamical structures with different associated time scales. In this regard, planetary atmospheric circulation can define (persistent) teleconnection processes, which influence, for example, the increase in the surface temperature of the Pacific Ocean that affects the meteorological circulations in different parts of the planet.

In South America, meteorological dynamics are directly influenced by the circulations occurring in both the Pacific and the Atlantic Oceans. In the Pacific, the main feature of circulation, on a global scale, is a vertical movement along the equator (west–east), defined as the Walker circulation, associated with ENSO events, which causes changes on a planetary scale. Another planetary circulation structure that extends from the equator to the latitudes of approximately 30 degrees in both hemispheres is Hadley circulation. These two planetary circulations act together, creating a complex structure. When there are lower temperatures, below the climatological average, on North Atlantic Ocean, near the Caribbean, at the Brazilian coast there are higher average temperatures on the sea and excessive rainfall at Brazilian Northeast, with the intensification of the Intertropical Convergence Zone (ITCZ). This situation is known as the Atlantic temperature gradient, associated with Hadley circulation. SST in Equatorial Atlantic is subject to interannual variability that has impacts on the surrounding continents. (Lübbecke *et al.* 2018; Dippe *et al.* 2018; Scaife *et al.* 2018).

The results show that these equatorial surface anomalies are responsible for the seasonal migration to the North of the ITCZ in early summer in Brazilian North and Northeast. Ferreira and Mello (2005) showed that the atmospheric circulation over the tropical region is strongly modulated and modified by thermodynamic patterns on Pacific and Atlantic basins. Thus, in the years when anomalies of positive or negative SST occur in the basins of these oceans, the Hadley cell, which acts southward (ascending branch in the tropics and descendant branches in the middle latitudes), and the Walker cell, zonal (ascending branch in the Western Pacific and descending branch in the Eastern Pacific), are disturbed, causing strong anomalies in the atmospheric circulation on the tropics. Consequently, these cells are displaced from their climatological positions.

A work by and Kossin (2007) showed the possible association between that AMM Index and hurricane activity, the dynamical "mode" of variability intrinsic to the tropical coupled ocean–atmosphere system, is strongly related to seasonal hurricane activity on both decadal and interannual timescales.

For the tropical Atlantic region, an index was created that represents this variability of the Atlantic temperature gradient: AMM SST index data were obtained from National Oceanic and Atmospheric Administration (NOAA; NOAA Earth System Research Laboratory – Physical Sciences Division; https://www.esrl.noaa.gov/psd/data/timeseries/monthly/AMM/), (Chiang and Vimont 2004; Kossin and Vimont 2007; Vimont and Kossin 2007).

This AMM SST index, to the tropical Atlantic region, allows performing cross-wavelet and wavelet analyzes to correlate it with the sunspot activity and analyzing multidecadal cycles in the respective series. As the historical AMM index series, it is a measure of the dynamic behavior associated with the interhemispheric gradient of the SST anomaly, which occurs over the tropical Atlantic Ocean region, between Brazilian Northeast/North and the coast of Africa. This work aims to analyze the variability of sunspots and their association to AMM Index, as well as to investigate the existence of multidecadal variabilities and their influence on the surface temperatures in Brazilian North and Northeast.

## METHODOLOGY

### DATA ANALYZED

*Monthly Climate Time series*

AMM SST Index data is defined via applying Maximum Covariance Analysis (MCA) to SST and the zonal and meridional components of the 10 m wind field over the time period of 1950 to 2005, from the National Centers for Environmental Prediction (NCEP)/National Center for Atmospheric Research (NCAR) Reanalysis. To define the spatial pattern, it is needed to also define the over the region (21S-32N, 74W-15E), and spatially smooth it (three longitude by two latitude points). The seasonal cycle is removed, data are detrended, a three-month running mean is applied to the data, and the linear fit to the Cold Tongue Index (CTI), a measure of El Niño Southern Oscillation (ENSO) variability (Zhang et al. 1997), is subtracted from each spatial point. Spatial patterns are defined as the first left (SST) and right (winds) maps resulting from the singular value decomposition of the covariance matrix between the two fields. AMM time series is calculated by projecting SST or the 10 m wind field (detrended, CTI removed, but no 3-month running mean) onto the spatial structure resulting from MCA. Significant monthly average data on the sunspots number (SN) were obtained at http://www.sidc.be/silso/datafiles (data from the Sunspot of the World Data Center [Sunspot Index and Long-term Solar Observations; SILSO], Royal Observatory of Belgium, Brussels), transferring the data file SN_m_tot_V2.csv with information from 1749 to 2018, with a 267 years series. The analyses of the time series were performed using the information obtained in the two aforementioned sources.

*Wavelet Analysis*

The WaveletComp is an R package used for analysis based on uni- and bivariate time series wavelets. Its 1.0 version (R Foundation for Statistical Computing) (Roesch and Schmidbauer 2014), applies frequency analysis of uni- and bivariate time series using Morlet *et al.* (1982a, 1982b; Goupillaud *et al.*1984), as well as the biwavelet package by Grinsted *et al.* (2004). Morlet wavelet, in the version implemented in WaveletComp, is defined as the convolution of the series with a set of "daughters wavelet" generated by the "mother wavelet", using time translation by τ and defining the scale by *s*. The position of the daughter wavelet in the time domain is determined by the location of the time parameter τ being displaced by *dt*, a time increment. WaveletComp rectifies the wavelet power spectrum (cross-wavelet) according to Liu *et al.* (2007), in the univariate case, and Veleda *et al.* (2012), in the bivariate case, to avoid "biased" results in filtering and estimating the associated high frequencies or in the sense of variability estimates with short periods in time series. The analyzed phenomena and the time series would tend to be underestimated by conventional approaches. Implemented options for smoothing in a period of time and/or a direction are necessary to perform the calculation with the coherence wavelets methodology, with their multiples (Liu, 1994). The set choice of scales *s* determines the series wavelet coverage in the frequency domain.

*Surface Weather Data*

The monthly temperature data were used from five meteorological stations of the Brazilian Air Space Control Department (DECEA; in Portuguese, Departamento de Controle do Espaço Aéreo) climate database. Minimum, average and maximum monthly mean temperature data were used at the airports of Belém, São Luiz, Fortaleza, Natal, and Fernando de Noronha. The time series were analyzed from January 1951 to September 2017, a period of 67 years. The anomalies were calculated in relation to the observed mean value of each series, in which the anomaly series of the monthly average of sunspots, and the minimum, average and maximum monthly temperatures were generated. ERA5 model reanalysis data is also used to characterize the AMM index modes. This is the reanalysis of the European Center for Medium-Range Weather Forecasts (ECMWF) – ERA5 reanalysis data. This meteorological dataset, ERA5 (Hersbach and Dee 2016), provides estimates of atmospheric parameters (such as air temperature, pressure, wind, humidity, and ozone at different altitudes) and surface parameters (such as rainfall, soil moisture, and sea-surface temperature), all at a resolution of about 31 km worldwide, and information on wave height over the global oceans. More information about the ERA5 model can be found at https://confluence.ecmwf.int//display/CKB/ERA5+data+documentation.

*Non-parametric Permutation Test*

A non-parametric statistical test was used to evaluate the correlation of the time series. The test used employs a statistical permutation (Collingridge 2013; Konietschke and Pauly 2014; Koopman *et al.* 2015; Zhang *et al.* 2017; Derrick *et al.* 2017; Pauly *et al.* 2018). As a definition, the vectors *P*, with the monthly values of the series anomaly in relation to their monthly average temperature, and *J* (N × 1), with the observed monthly average sunspot, intend to make random permutations of *J*, keeping *P* fixed. For each permutation, it calculated the correlation between vectors *P* and *J*, resampling the series in the order of 10,000 times, thus building the distribution of correlations (*r*). From these distributions, the test can obtain the value that represents the confidence interval at the level of 5% of the correlations in the upper or lower tails of distribution (*r* critical). The time series has analyzed 801 values. It used a subroutine program in Matlab©.

The permutation test method is more robust statistically to test the correlation signal between different series that have low correlation values. The correlation between the two series will have a positive upper tail and a negative lower tail, in 10,000 times. Exchanging the same series at random, the probability distribution can be reconstructed in order to analyze the test using a command that orders the randomly calculated correlations. In 10,000 times, the correlation in 500$^{th}$ position and the one in 9,500$^{th}$ position would be critical values at the level of 5%, because, if the original series has a correlation value greater than the value obtained at the 9,500 positions, and/or less than the value in the 500$^{th}$ position, it could not be said that the original correlation is random. The lower critical value of the tail, in the 500$^{th}$ position, was estimated by the series of the minimum average monthly temperature anomaly. However, the value of the upper tail, 9,500$^{th}$ position, was estimated by the average maximum monthly temperature anomaly. It can be concluded that the series must have some degree of significant correlation. There is statistical evidence that the observed correlations are significant and cannot be discarded. If the correlation falls between the critical values, nothing can be said, which characterizes randomness.

# RESULTS

Figures 1a and 1b represent the wind intensity in December 2017 (a) (with a positive AMM index value of 6.14 in SST and a negative value of –2.39 in wind) and in June 2018 (b), with a negative value of –6.48 in SST and a positive value of 4.39 in wind. These figures show two modes of the AMM index in antagonistic positions. Such positions have a dynamic implication, affecting the circulation and characterizing the position of the ITCZ, which affects seasonal meteorological behavior in the North and Northeast regions of Brazil. Figure 1a shows a not well-defined position of the ITCZ, and presents a greater intensity of the wind along the Brazilian coast. However, in Fig. 1b (10ºS –15ºS), it is displayed a greater intensity southward, in the center of the Atlantic Ocean, in the Southern hemisphere, and in the coast of Africa. It also presents a greater intensity of the wind southward in the Northern hemisphere, creating a significant gradient, and positioning the ITCZ between the Brazilian Northeast and the African continent, in 5ºN.

Figures 1c and 1d show the mean temperature over the tropical Atlantic Ocean, having in ºC a range with higher temperatures in the environment at 15°N above 24 °C. Figure 1d shows lower average temperatures in the order of 20 to 22 °C at 15°N on the coast of the African continent. In South Atlantic, the temperature in 10ºS is higher, ranging between 22 to 24 ºC and 24 to 26 ºC. Figs. 1e and 1f show the total precipitation. Figure 1e depicts a strip in the coast of Africa at 5°N and 5°S, where low convective activity and lower values of total precipitation are seen. Figure 1f shows a well-defined range of precipitation at 5°N, between American and African continents.

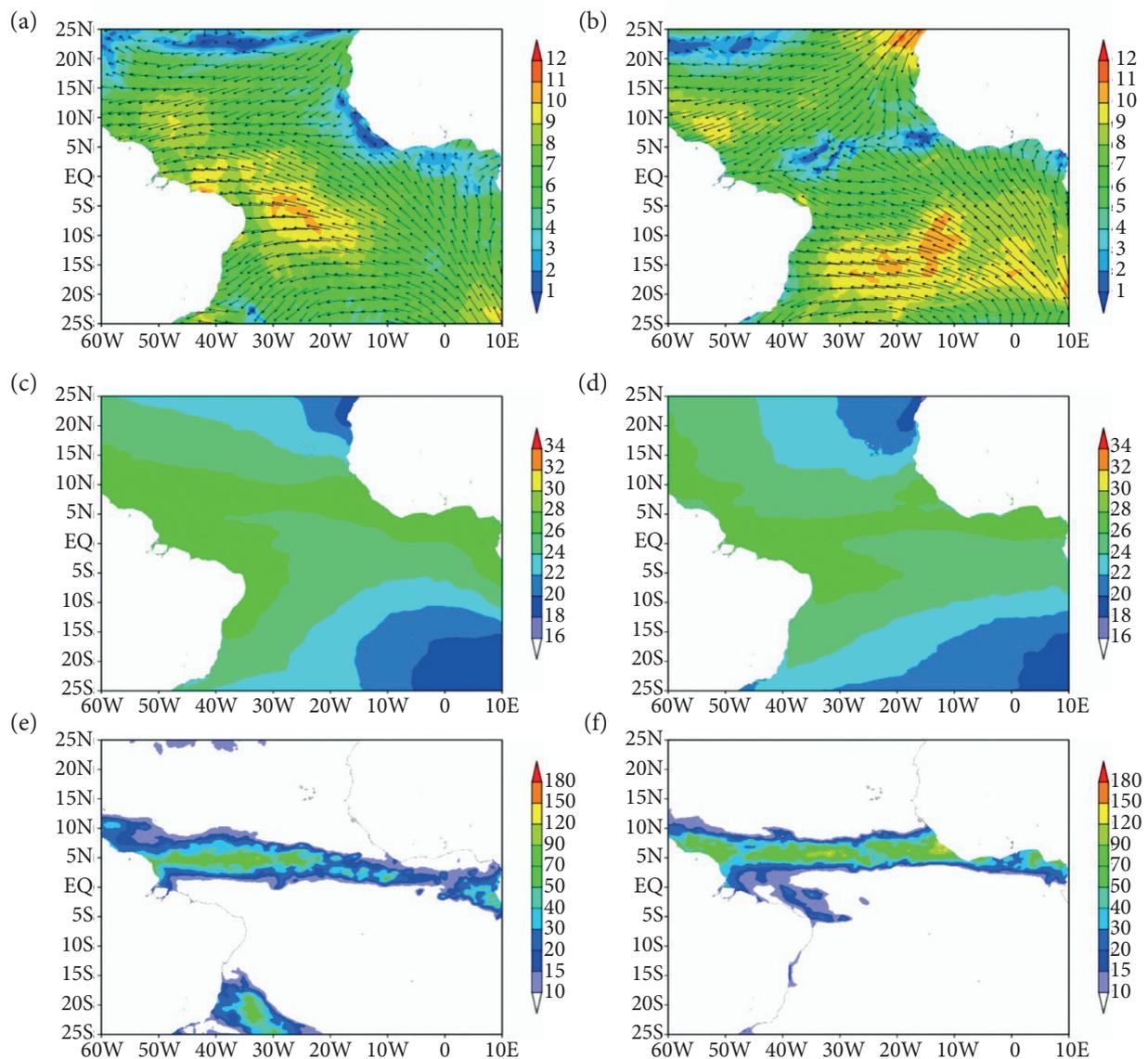

**Figure 1.** Reanalysis information of the ERA5 model in December 2017 (a, c and e) and June 2018 (b, d and f), where a and b are the monthly wind averages in ms$^{-1}$. Figures c and d represent the monthly average temperature in °C. Figures e and f are monthly averages of total precipitation in mm.

The behavior of the AMM Index represents the dynamic interaction between the interhemispheric in the tropical Atlantic Ocean, modulating the displacement of the ITCZ, and showing the coupling between the ocean and the atmosphere with their different temporal scales involved. Figure 2 shows in the series the AMM SST Index, in Fig. 2-A presents cycles with periods of 12 months, 32 months, 64 months and 128 months and a weak signal in 256 months. In Fig. 2-B in sunspot appear 32 months, 64 months and 128 months and 256 months, corresponding to 2.66 years, 5.33 years, 10.66 years, and 21.33 years. In Fig. 2-C in wind appear 32 months, 64 months and 128 months, corresponding to 2.66 years, 5.33 years and 10.66 years. The AMM Index-Wind appears predominantly in periods less than 16 months. What the analysis allows to conclude that shows the Wind Field is associated more significantly with shorter time scales of the order of 32 months and less than 16 months, but presents a remarkable signal in 128 months, approximately eleven years.

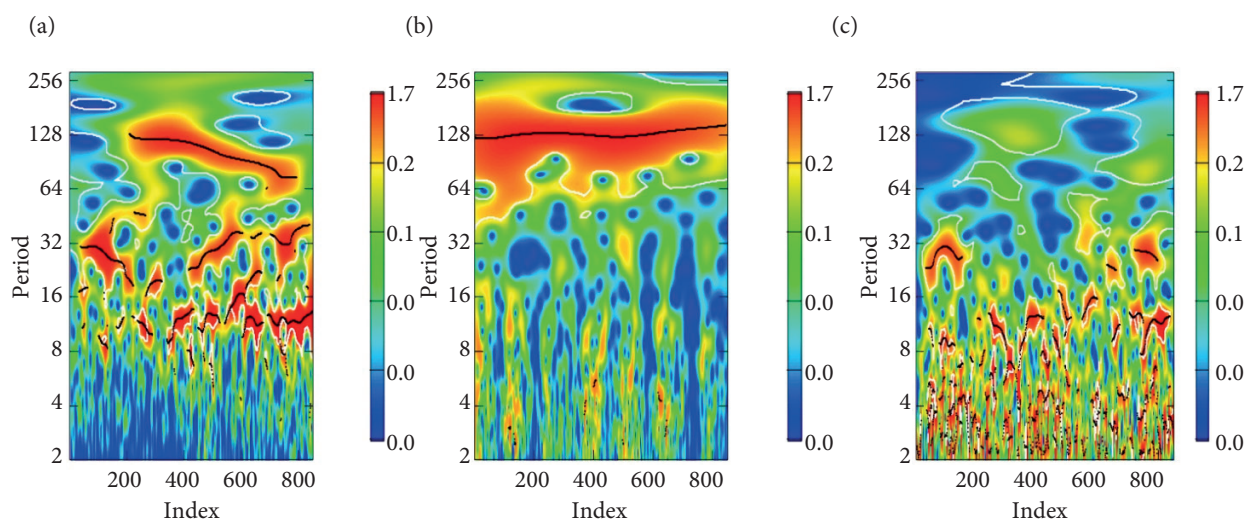

**Figure 2.** Wavelet analysis of AMM index time series. (a) SST; (b) sunspot; (c) wind.

The analysis of cross-wavelets in Fig. 3 shows that the 32-months period may be associated with the quasi-periodic oscillations of equatorial zonal wind between the easter- and westerlies, known as the quasi-biennial oscillation (QBO), in the lower stratosphere, and the tropospheric biennial oscillation (TBO), in the troposphere. Both have similar oscillation features (Kwan and Samah 2003; McCormack 2003) and have great eleven years influence in the analysis of cross-wavelet by the AMM index (AMM SST index [a] and AMM wind index [b]). In Figs. 3a and 3c, a significant signal in 21.33 years is seen, which does not appear in the same way in the AMM wind index (b and d), compared to sunspots.

Table 1 presents the results of the correlations observed between the series of anomalies in the minimum, average and maximum monthly temperatures and the series of monthly sunspot anomalies. The non-parametric test method is statistically more robust than the classical test method and does not present the problems of statistical inference for its analysis and results. Table 1 shows that the maximum temperature anomalies are correlated with the sunspot anomalies in Belém, São Luiz, Fortaleza, and Fernando de Noronha. The original correlation value is greater than the one observed in the permutation test. There is statistical evidence, at 5% level, that the series of anomalies are correlated. It cannot be ruled out that the original correlation is not random in the test of 10,000 permutations of the original series. The historical series with the highest correlation observed was in the surface station of São Luiz, said correlation being of 0.18 with the anomalies of the sunspots. However, all series of the analyzed minimum temperature anomalies showed a negative correlation when it was observed the minimum monthly average temperature anomaly, except for São Luiz, characterizing that, when the minimum temperature anomaly occurs, the maximum of the sunspot anomaly did not.

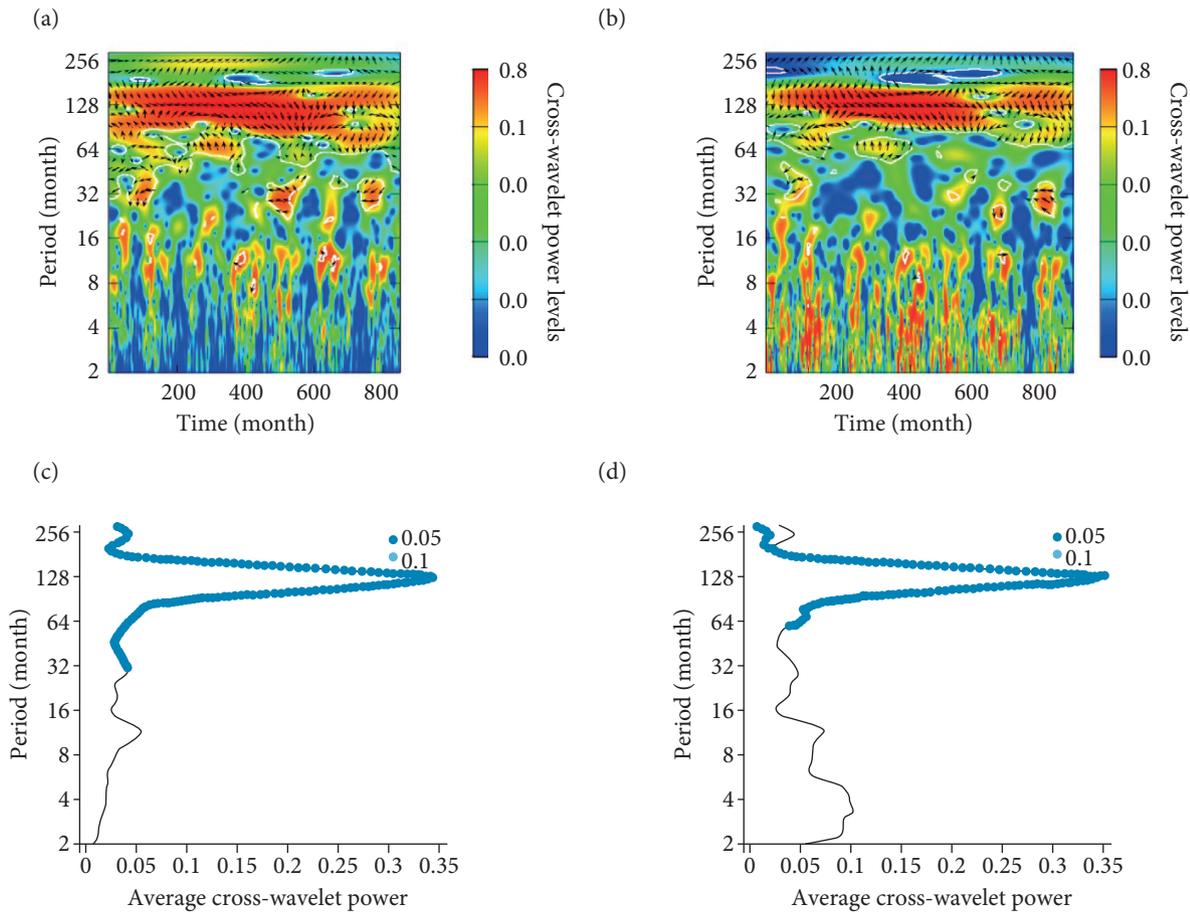

**Figure 3.** The bivariable energy spectrum with crossed wavelet: (a) AMM – SST and sunspot; (b) AMM – wind and sunspot. It also presents the bivariable cross-wave mean time spectrum: (c) AMM – SST and sunspot; (d) AMM – wind and sunspot.

**Table 1.** Result of the correlations between temperature anomalies and sunspots with the non-parametric test.

| City | Correlation of temperature/sunspots monthly anomalies | | | | |
|---|---|---|---|---|---|
| | Minimum | Mean | Maximum | Non-parametric test | |
| | | | | Lower tail | Upper tail |
| Belém | −0.07 | 0.01 | 0.09 | −0.058 | 0.059 |
| São Luiz | 0.01 | 0.09 | 0.18 | −0.057 | 0.058 |
| Fortaleza | −0.05 | 0.04 | 0.06 | −0.058 | 0.058 |
| Fernando de Noronha | −0.16 | −0.17 | 0.07 | −0.060 | 0.060 |
| Natal | −0.04 | −0.03 | 0.01 | −0.057 | 0.058 |

N = 801 values.

Figure 4 shows the distributions of correlations generated by the random permutation process of the original correlation between the series of maximum temperature and sunspot anomaly. The non-parametric correlation test is a statistically robust method, because it showed that, at the 5% level, the original correlations (above the critical value of the 500$^{th}$ position, in the lower tail, and the 9,500$^{th}$, in the upper tail, in 10,000 permutations) could not be discarded.

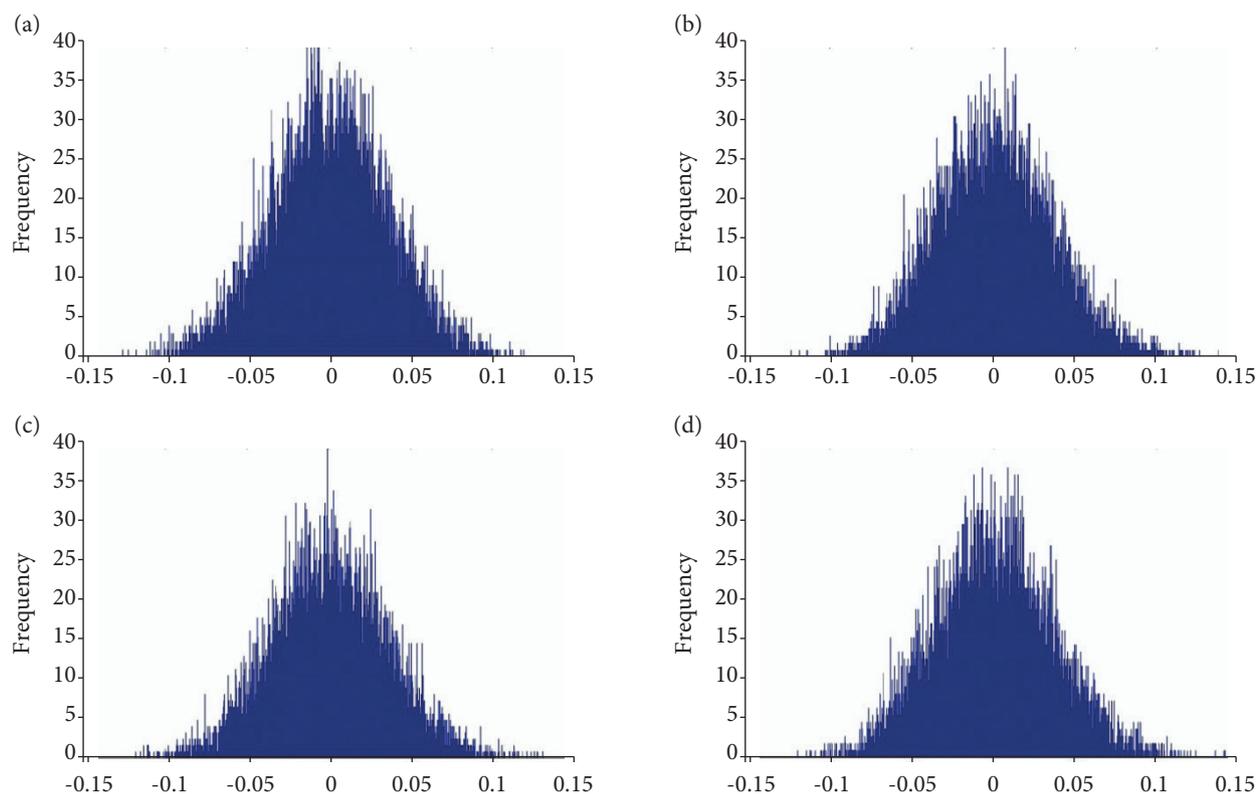

**Figure 4.** Distributions of correlations generated by the non-parametric test of the series of maximum temperature anomalies correlated with the sunspots anomaly. (a) Belém; (b) São Luiz; (c) Fortaleza; (d) Fernando de Noronha.

## CONCLUSIONS

The AMM index consists of two parts, one associated with the SST and the other to the wind field over the tropical Atlantic region. In this aspect, the results of this work show that an association between the series of the AMM index and the sunspot time series, through the analysis of cross-wavelet, presented the existence of multidecadal cycles. In the SST index series, a well-defined eleven-year signal was found, as well as one of 21.33 years. However, in the wind section of this index, the cycle of 21.33 years did not appear to be significant, showing instead a tendency to represent periods of less than 32 months. It also presented a signal of 128 months (cycle of eleven years).

The solar activity directly affects the terrestrial magnetosphere, creating electrodynamic and hydrodynamic effects in the upper atmosphere by the solar wind (X-rays, high-energy particles, solar plasma, etc) (Kane 2005; Lundin *et al.* 2007; Yigit *et al.* 2016). These characteristics act in the stratosphere and the troposphere, affecting the surfaces of the oceans, in the value observed in their SST, and generating circulations with different temporal and spatial scales, with complex and turbulent processes. The solar activities characterize persistent cycles, with dominant frequencies, affecting significantly the terrestrial surface, and the atmosphere/ocean coupling, which reflects in planetary level, with a certain degree of memory. Therefore, the analyzes of minimum, average, and maximum temperatures in five locations of Brazilian North and Northeast showed that they accompany, with some degree of synchronism, and are correlated to the positive anomalies of the sunspots, with the maximum surface temperature (Belém, São Luiz, Fortaleza, and Fernando de Noronha). The non-parametric test of correlation was able to show a certain degree of association between the historical series observed (1951 to 2017). This work shows a new possibility for future studies with high troposphere and stratosphere data analysis that characterize solar wind activity, associated with the multidecadal cycles of solar activity

## AUTHOR'S CONTRIBUTION

Conceptualization, Correa CS; Guedes RL and Corrêa KAB; Formal analysis, Correa CS; Guedes RL; da Rocha AMM and Corrêa KAB; Supervision, Correa CS; Guedes RL; da Rocha AMM and Corrêa KAB; Writing – Original Draft, Correa CS; da Rocha AMM and Corrêa KAB; Writing – Review & Editing, Correa CS and Corrêa KAB.